\documentclass[preprint]{aastex}
\usepackage{graphicx,amssymb,amsmath,times,verbatim}
\usepackage{color}

\usepackage{amssymb,amsmath}
\usepackage[normalem]{ulem}

\def\lta{\mathrel{\spose{\lower 3pt\hbox{$\mathchar"218$}}
     \raise 2.0pt\hbox{$\mathchar"13C$}}}
\def\gta{\mathrel{\spose{\lower 3pt\hbox{$\mathchar"218$}}
     \raise 2.0pt\hbox{$\mathchar"13E$}}}

\def\mathnew{\mathsurround=0pt}

\def\simov#1#2{\lower .5pt\vbox{\baselineskip0pt \lineskip-.5pt
\ialign{$\mathnew#1\hfil##\hfil$\crcr#2\crcr\sim\crcr}}}

\shorttitle{$\gamma-$ray Variability of PKS 1222+216}
\shortauthors{Kushwaha, Singh \& Sahayanathan}

\begin{document}


\title{Brightest \emph{Fermi}-LAT Flares of PKS 1222+216: Implications on Emission and Acceleration Processes}
\author{Pankaj Kushwaha\altaffilmark{1}, K. P. Singh\altaffilmark{1}, Sunder Sahayanathan\altaffilmark{2}}
\altaffiltext{1}{Department of Astronomy \& Astrophysics, Tata Institute of Fundamental Research, Mumbai, India; 
pankaj563@tifr.res.in}
\altaffiltext{2}{Astrophysical Sciences Division, Bhabha Atomic Research Centre, Mumbai, India}
 

\begin{abstract}
We present a high time resolution study of the two brightest $\gamma$-ray outbursts
from a blazar PKS 1222+216 observed by the \textit{Fermi} Large Area
Telescope (LAT) in 2010. The $\gamma$-ray light-curves obtained in four different
energy bands: 0.1--3, 0.1--0.3, 0.3--1 and 1--3 GeV, with time bin of 6 hr,
show asymmetric profiles with a similar rise time in all the bands but
a rapid decline during the April flare and a gradual one during the June.
The light-curves during the April flare show $\sim 2$ days long plateau in
0.1--0.3 GeV emission, erratic variations in 0.3--1 GeV emission, and a daily
recurring feature in 1--3 GeV emission until the rapid rise and decline within
a day. The June flare shows a monotonic rise until the peak, followed by a
gradual decline powered mainly by the multi-peak 0.1--0.3 GeV emission. The
peak fluxes during both the flares are similar except in the 1--3 GeV band
in April which is twice the corresponding flux during the June flare. Hardness
ratios during the April flare indicate spectral hardening in the rising phase 
followed by softening during the decay. We attribute this behavior to the
development of a shock associated with an increase in acceleration efficiency
followed by its decay leading to spectral softening. The June flare
suggests hardening during the rise followed by a complicated energy dependent behavior
during the decay. Observed features during the June flare favor multiple emission
regions while the overall flaring episode can be related to jet dynamics.
\end{abstract}

\keywords{radiation mechanisms: non-thermal -- galaxies: active -- FSRQs: individual: PKS 1222+216
(4C +21.35) -- galaxies: jets -- X-rays: galaxies}

\section{INTRODUCTION} \label{sec: intro}
The rapid, complex, high amplitude and broadband (radio to $\gamma$-rays) variability is a
defining characteristic of a class of radio-loud active galactic nuclei (AGN) called blazars.
These observed properties are believed to be a result of relativistic motion of non-thermal
plasma along the jet, oriented at small angles to the observer's line of sight \citep{1995PASP..107..803U}. The 
erratic variability combined with the complexity of organizing simultaneous multi-band
observations and limitations of modern telescopes to resolve the emitting region makes
multi-band variability studies and correlations as important tools to infer the underlying
physical processes and the environment of AGNs. A detailed and systematic study of temporal
and spectral features in different energy bands can offer a potential diagnostic for
understanding the radiation mechanisms and underlying physical processes \citep{1998A&A...333..452K,
2000ApJ...536..299K,2005NewA...11...17B}. For $\gamma$-ray bright AGNs with good photon statistics,
the broadband energy coverage and continuous operation of \textit{Fermi}-LAT offer a unique
opportunity to extract multi-waveband light-curves down to scanning timescale of $\sim 3$
hours (see \citet{2013APh....48...61V} for more details), making such a study feasible at
$\gamma$-ray energies.

The radiative output of high luminosity blazars, particularly the flat spectrum radio
quasars (FSRQs) is mostly dominated by $\gamma$-ray emission in the GeV regime and
beyond in some cases \citep{2008MNRAS.384L..19B}. This emission is generally explained
as a result of inverse Compton (IC) scattering of photons external to the jet by
non-thermal electrons \citep{2012MNRAS.425.2519N, 2013ApJ...779...68S}, commonly 
referred as external Compton (EC). Two sites-- one at sub-parsec scales and another at
parsec scales, have been suggested for dominant $\gamma$-ray emission/contribution
\citep{2013hsa7.conf..152A, 2012ApJ...758L..15D}. On sub-parsec scales ($<$ 1 pc),
$\gamma$-ray emission is due to the EC scattering of photons from broad line region (BLR) and/or
accretion-disk \citep{2010ApJ...714L.303F}, while on parsec scales the EC scattering of infrared
(IR) photons from the putative molecular/dusty torus can be dominant. The sub-parsec
scenario has been very successful in explaining the observed SED and short time $\gamma$-ray
variability of blazars until very recently. However, the latest radio and $\gamma$-ray
correlation studies \citep{2013hsa7.conf..152A,2014MNRAS.441.1899F} favor parsec
scenario in most cases with further support from the detection of FSRQs at very high
energies \citep[VHE;][]{2008Sci...320.1752M, 2010HEAD...11.2706W,2011ApJ...730L...8A}. 
In the parsec scale scenario; however,  particle acceleration, bulk velocity, and
emission regions associated with short time variability of few minutes to hours are
matters of ongoing debate due to its larger distance from the central engine
\citep{2012MNRAS.425.2519N, 2013hsa7.conf..152A, 2014MNRAS.442..131K}. 

Most of the studies till date have relied on spectral features \citep{2011ApJ...733L..26A,
2011ApJ...733...19T} and constraints from $\gamma$-ray opacity to infer the underlying
mechanisms, seed photons and hence the location of emission sites in blazars
\citep{2011ApJ...730L...8A, 2011ApJ...733...19T, 2012MNRAS.425.2519N, 2012MNRAS.419.1660S,
2014MNRAS.442..131K}. However, spectral information from $\gamma$-ray emission is subject to
Klein-Nishina (KN) softening, pair production and presence of intrinsic features in the particle
spectra,  making GeV emission and underlying processes hard to disentangle \citep{2012ApJ...758L..15D}.
Moreover, if the underlying physical processes (particle acceleration and injection) are evolving
faster than the cooling timescales, as suggested by the observations in cases of high luminosity
blazars \citep{2014MNRAS.442..131K}, spectral information averaged over a longer duration data may
lead to false inferences regarding these processes. High time resolution study of spectral, timing and 
variability features though, can be used to infer these timescales and their relative dominance, at least
for bright sources. This, in fact, can also differentiate between multiple injections/emission-regions
which on larger timescales seem coherent otherwise \citep{2013ApJ...766L..11S, 2013MNRAS.431..824B}. However, only
a truly simultaneous multi-band study holds the key to gain further insight into these enigmatic processes
\citep{1998A&A...333..452K,1999MNRAS.306..551C}.

PKS 1222+216 (4C +21.35\footnote{PKS 1222+21, PKS B1222+216}; $z=0.432$) is a 
lobe dominated FSRQ at long radio wavelengths (cm onwards) and a known
$\gamma$-ray emitter \citep[and references
therein]{2011ApJ...733...19T,2014ApJ...786..157A}. In high resolution radio maps, it shows
a strong jet bending \citep{1993AJ....105.1658S, 2007ApJS..171..376C} and superluminal 
knots with complex 3D trajectories suggesting non-radial accelerations \citep{2012IJMPS...8..163H, 
2014ApJ...786..157A}. Increased $\gamma$-ray activity from the source was
reported by \emph{Fermi}-LAT during April 2009 \citep{2009ATel.2021....1L}.
Since December 2009 till mid 2010, it has been particularly active at $\gamma$-ray
energies undergoing frequent and rapid flux variations \citep{2009ATel.2349....1C}.
Enhancements were also seen by other observatories operating at similar/different
energy-bands \citep{2009ATel.2348....1V,2010ATel.2641....1B,2010ATel.2626....1C}.
During this period the source showed two prominent $\gamma$-ray flares ($>10^{-5}$ph
cm$^{-2}$ s$^{-1}$): one in April 2010 \citep{2010ATel.2584....1D} and another in
June 2010 \citep{2010ATel.2687....1I}, the brightest so far in LAT band from this
source\footnote{http ://fermi.gsfc.nasa.gov/FTP/glast/data/lat/catalogs/asp/current/lightcurves/PKSB1222+216\_86400.png}.
A detailed investigation of LAT data during April flare confirmed VHE emission
\citep[129 GeV photon;][]{2010ATel.2617....1N,2011A&A...529A..59N} leading to the
inclusion of PKS 1222+216 in the LAT VHE catalog \citep{2011A&A...529A..59N}. During
this high $\gamma$-ray activity period, two VHE excesses were detected by the \emph{MAGIC}
observatory-- one immediately after the end of the April flare on MJD 55319.97
\citep{2014ApJ...786..157A} and other one during the rising part of the June
2010 flare \citep{2010ATel.2684....1M,2011ApJ...730L...8A} establishing PKS 1222+216
as a potential VHE FSRQ. 

First two years of \emph{Fermi}-LAT observations of PKS 1222 + 216 have been analyzed by
\citet{2011ApJ...733...19T} by dividing the LAT light-curve in different activity states
based on the observed $\gamma$-ray variability of the source. They have further performed
a detailed investigation of the LAT data for the time period considered in this paper (the
\emph{Active} state of \citet{2011ApJ...733...19T}), focusing particularly on the spectral
evolution, $\gamma$-ray emission and energetics of the source during its different activity
states. Based on the spectral fit of LAT spectral energy distribution (SED) and the stability
of the break energy during its various activity states, the authors argued that the break can
be well explained by recombination in the BLR as suggested by \citet{2010ApJ...717L.118P}. However,
EC scattering of BLR photons (Lyman-$\alpha$) to VHE lies in the KN regime, making VHE emission extremely
inefficient with a significant emission only up to few 10s of GeV \citep{2012MNRAS.425.2519N,
2014MNRAS.442..131K}. Detection of VHE photons during both the prominent flares in 2010 and a hard VHE spectra during
the June 2010 flare \citep{2011ApJ...730L...8A} contradict the BLR origin \citep{2011A&A...534A..86T,
2012MNRAS.419.1660S,2012MNRAS.425.2519N}, though it can contribute significantly at GeV energies.

In this paper, we present a detailed systematic analysis of variability during the two
brightest $\gamma$-ray flares of PKS 1222+216 mentioned above. The continuous broadband
observation of \textit{Fermi}-LAT and the strong fluence allowed us to extract $\gamma$-ray
light-curves down to a timescale of 6 hours in four different LAT energy bands: 0.1--3 GeV, 0.1--0.3 GeV,
0.3--1 GeV and 1--3 GeV. Details of \textit{Fermi}-LAT data and its reduction procedures
are discussed in Section \ref{sec:data}. In Section \ref{sec:result}, we present the main results
obtained from the LAT analysis, with its implications on blazar emission and underlying physical processes
being discussed in Section \ref{sec:discussion}. Finally we summarize our finding in Section
\ref{sec:conclude}. A standard $\Lambda$CDM cosmology has been assumed throughout this paper
with $\Omega_m = 0.3$, $\Omega_\Lambda = 0.7$ and $H_0 = 70$ km s$^{-1}$ Mpc$^{-1}$.

\section{DATA REDUCTION}\label{sec:data}

The pair conversion detector known as Large Area Telescope (LAT) on board \emph{Fermi
Gamma-ray Space Telescope} is sensitive to $\gamma$-ray photons with energies $20$~MeV
to $>300$~GeV \citep{2009ApJ...697.1071A}. It operates mostly in survey mode scanning
the entire sky every $\sim 3$ hours, thereby offering a unique capability for
studying short time evolution of $\gamma$-ray sources. Here, we have analyzed the LAT
data of PKS 1222+216 from April-June 2010 (see Table \ref{tab:ObsLog}) when it was
undergoing intense $\gamma$-ray activity and had emitted two brightest GeV
flares of intensity $\gtrsim 10^{-5}$ ph cm$^{-2}$ s$^{-1}$ with $\gamma$-ray
emission extending up to VHEs.

\begin{table}[ht]
\centering
\caption{Log of \textit{Fermi}-LAT observations}
\begin{tabular}{c c c}
\hline \hline
   Start Date (MJD)	&	End Date (MJD)	&  State 	\\
\hline
April 14, 2010 (55300)	&	June 27, 2010 (55375)	&	Active	\\
April 27, 2010 (55313)	&	May 01, 2010 (55318)	&  {\it Flare 1}	\\
June 16, 2010 (55363)	&	June 22, 2010 (55370)	&  {\it Flare 2}	\\
\hline 
\end{tabular}
\label{tab:ObsLog}
\end{table}

LAT data studied here (see Table \ref{tab:ObsLog}) were analyzed using \emph{Fermi}-LAT
Science tool version v9r27p1 following the recommended analysis
procedure\footnote{http://fermi.gsfc.nasa.gov/ssc/data/analysis/scitools/python\_tutorial.html}.
Only events classified as ``evclass=2'', energy $> 100$ MeV and zenith angles 
$<$ 100$^\circ$ from a region of interest (ROI) of $15^\circ$ centered around
the source were considered during our analysis to avoid calibration uncertainties
and the earth's limb. Good time intervals associated with the selected events were
calculated using the recommended selection ``(DATA\_QUAL==1)\&\&(LAT\_CONFIG==1)
\&\&ABS(ROCK\_ANGLE)$<52$''. Effects of various cuts on data and sources outside 
the ROI were taken into account by calculating exposure map on ROI and an additional
annulus of $10^\circ$ around ROI. The selected events were then analyzed using ``unbinned
maximum likelihood'' method  \citep[PYTHON implementation of \textit{gtlike}]{1996ApJ...461..396M}.
Sources within the angular field of exposure map were selected and modeled using
2nd LAT catalog \citep[2FGL -- gll\_psc\_v08.fit;][]{2012ApJS..199...31N} and the
\textit{Pass 7} instrument response function (\emph{P7SOURCE\_V6}). Galactic diffuse
emission and isotropic background were taken into account by using the respective
template (gal\_2yearp7v6\_v0.fits, iso\_p7v6source.txt).

The daily 0.1--300 GeV LAT $\gamma$-ray light-curve during the entire episode was extracted
following the above procedures and assuming a log-parabola model [$dN/dE\sim E^{-\alpha-\beta log(E)}$]
for the source \citep{2012ApJS..199...31N}. We have
used a logarithmically equi-spaced, twenty energy grids, per energy decade for the
spectral fit. Point sources with test statistic (TS) $\leq 0$ (equivalent of standard
deviation $\sigma^2$) were removed from the source model files during each analysis.
The 6 hr light-curves in four different LAT energy bands: 0.1 -- 3 GeV, 
0.1 -- 0.3 GeV, 0.3 -- 1 GeV and 1 -- 3 GeV, during \textit{Flare 1} and \textit{Flare 2}
were extracted using the best fit model files from the longer duration analysis as
input model. A TS value of $10 ~(\sim3\sigma)$ was used as a detection criteria for
the source. The systematic uncertainties associated with the derived fluxes are 10\%
at 0.1 GeV and 5\% between 0.316 GeV to 10 GeV.

\section{ANALYSIS AND RESULTS}\label{sec:result}

 \begin{figure}[ht]
 \centering
\begin{minipage}[t]{0.48\linewidth}
\includegraphics[scale=0.65,angle=0]{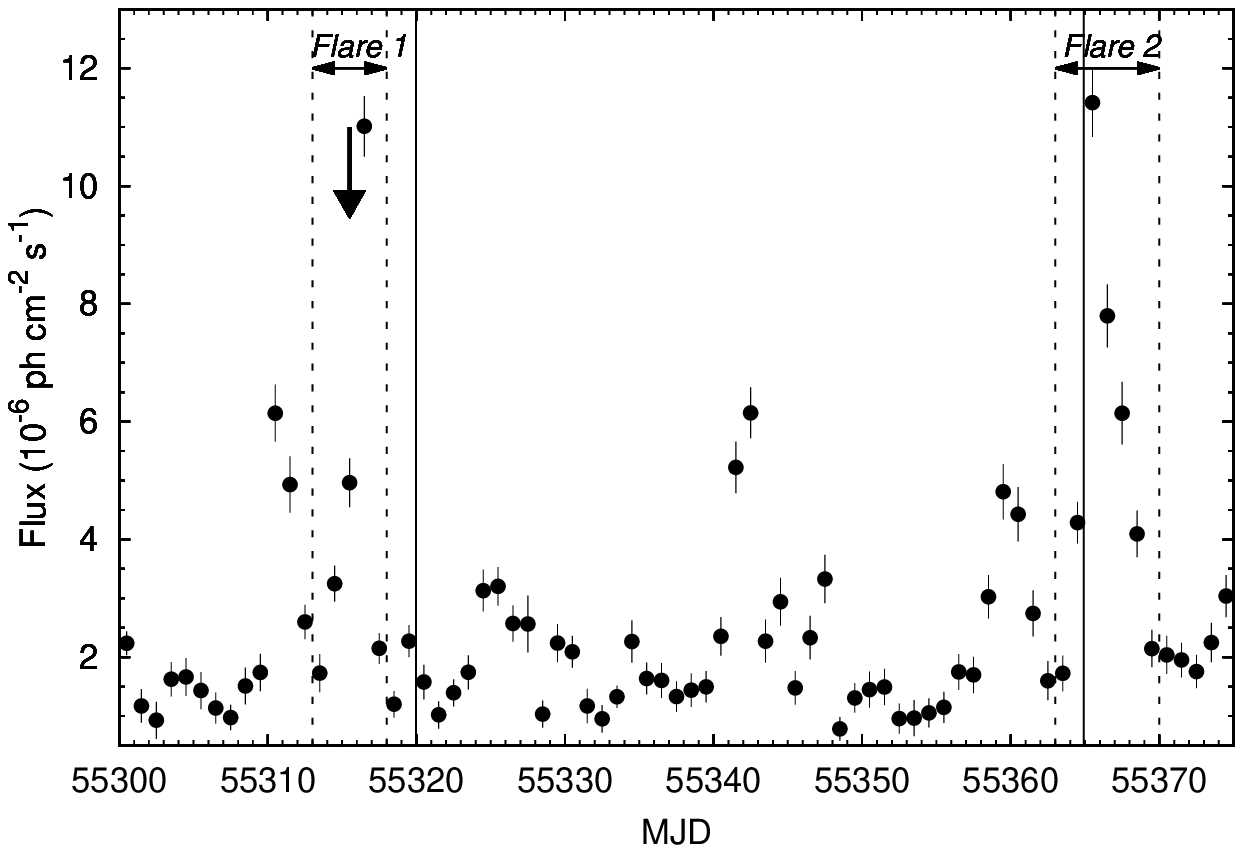}
\end{minipage}
 \quad
\centering
\begin{minipage}[t]{0.48\linewidth}
\includegraphics[scale=0.65,angle=0]{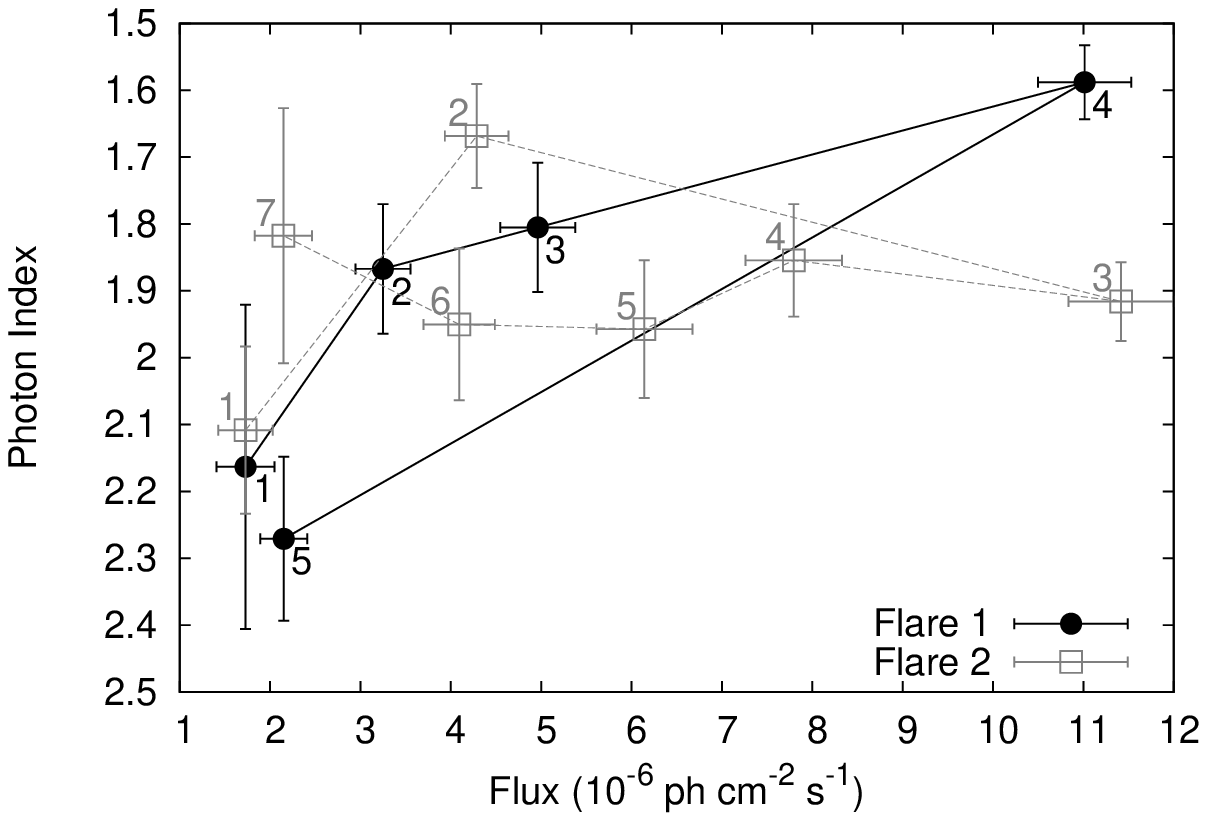}
\end{minipage}
\caption{\emph{Left}: The daily binned 0.1--300 GeV \textit{Fermi}-LAT light-curve of PKS 1222+216
during its active period (see Table \ref{tab:ObsLog}) in 2010. The dashed vertical lines mark the
intervals of the two brightest $\gamma$-ray flares from PKS 1222+216 observed by the \textit{Fermi}-LAT
till date. The vertical solid lines and the black arrow indicate the epochs of detection of VHE
emission by the \textit{MAGIC} observatory and the \textit{Fermi}-LAT respectively. Error bars
represent statistical 1$\sigma$ limit. \emph{Right}: Variation of spectral indices with observed
fluxes during \textit{Flare 1} (black solid line and filled circles) and \textit{Flare 2} (grey dashed
line and empty squares) marked by numbers indicating the day from the start of the respective flare.}
\label{fig:lcspec}
\end{figure}

The 0.1--300 GeV LAT $\gamma$-ray light-curve of the source extracted on a day timescale during
the time interval analyzed here (see Table \ref{tab:ObsLog}) is shown in the left panel of Figure
\ref{fig:lcspec}.  The period is characterized by a rapid and frequent flux variations at $\gamma$-ray
energies, and five major flares of fluxes $F>4 \times 10^{-6}$ ph cm$^{-2}$ s$^{-1}$ are
observed. Two of these, recorded respectively in April and June 2010 have similar amplitudes ($F
\gtrsim10^{-5}$ ph cm$^{-2}$ s$^{-1}$). The two flares, marked by dotted lines with
labels \textit{Flare 1} and \textit{Flare 2} respectively in the left panel of Fig. \ref{fig:lcspec}
are also the brightest $\gamma$-ray flares from this source in the history of \textit{Fermi}-LAT
operation till date. The right panel of Fig. \ref{fig:lcspec} shows the daily variation of
spectral index as a function of the observed flux (index-flux hysteresis) with numbers
indicating the day from the onset of the respective flare. A clear clockwise (CW) evolution
of index-flux hysteresis can be seen during both the flares. Interestingly, both the
flares were preceded by a relatively smaller amplitude ($\sim0.5$ of \textit{Flare 1} peak)
and similar duration flares. Despite their similar flux, CW evolution of index-flux
hysteresis, and the preceding history, both \textit{Flare 1} and \textit{Flare 2}
reveal a characteristically different temporal patterns during the flaring episodes. The
LAT flux during \textit{Flare 1} reaches its peak in 4 days and falls rapidly within a
day. On the contrary, \textit{Flare 2} reaches its peak within 3 days and shows a
gradual fall lasting 4 days. Gamma-ray emission extending up to VHE (\textit{Flare 1}:$\sim 130$ GeV,
\textit{Flare 2}:$\sim 400$ GeV) was detected just before the LAT peak during both these
flares. The vertical arrow in Fig. \ref{fig:lcspec} (left panel) shows the time when VHE
emission was detected in LAT during \textit{Flare 1} \citep[129 GeV,][]{2010ATel.2617....1N}
while the solid line during \textit{Flare 2} marks the VHE detection by the \textit{MAGIC}
observatory \citep{2010ATel.2684....1M, 2011ApJ...730L...8A}. Another $4.4\sigma$ VHE
detection has been claimed on MJD 55320 in the \textit{MAGIC} data, and is shown by a
vertical solid line in Fig. \ref{fig:lcspec} during the $\gamma$-ray active period of
the source \citep{2014ApJ...786..157A}.

\subsection{Flux Variability} \label{subsec:var}

\begin{figure}
\centering
\begin{minipage}[t]{0.485\linewidth}
\includegraphics[scale=1.,angle=0]{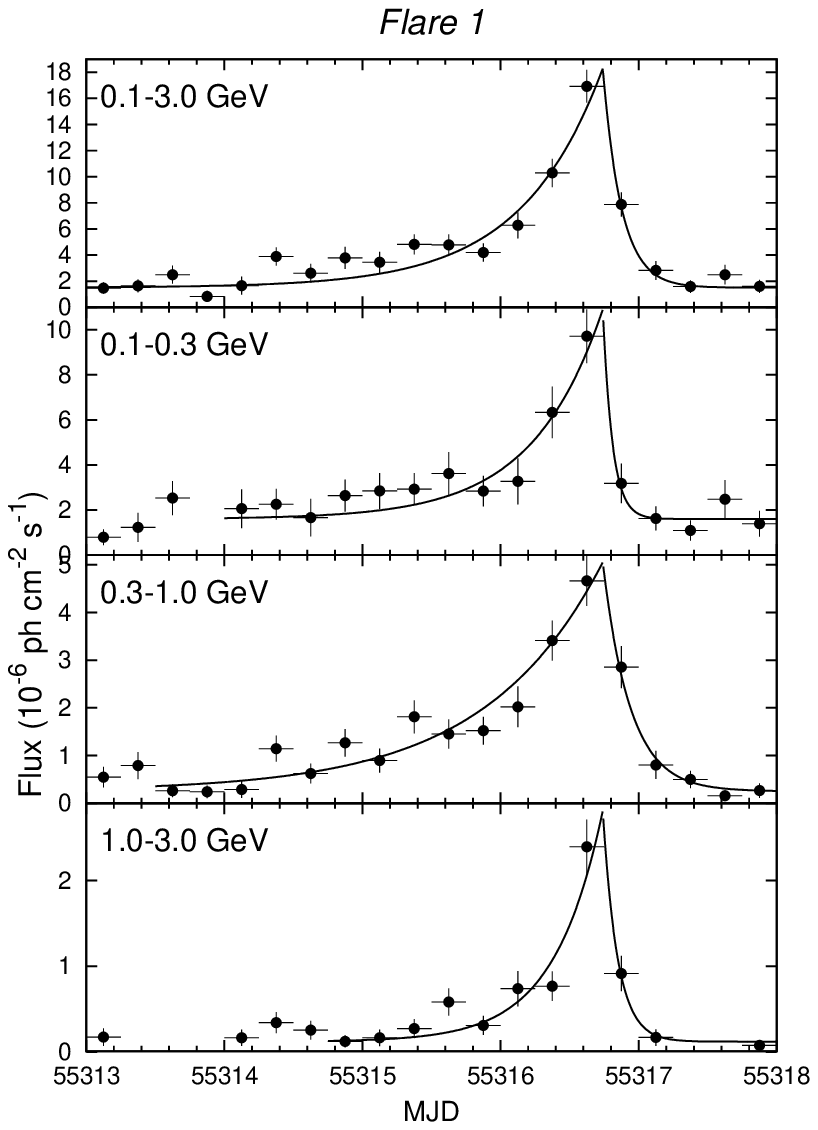}
\end{minipage}
 \quad
\centering
\begin{minipage}[t]{0.485\linewidth}
\includegraphics[scale=1.0,angle=0]{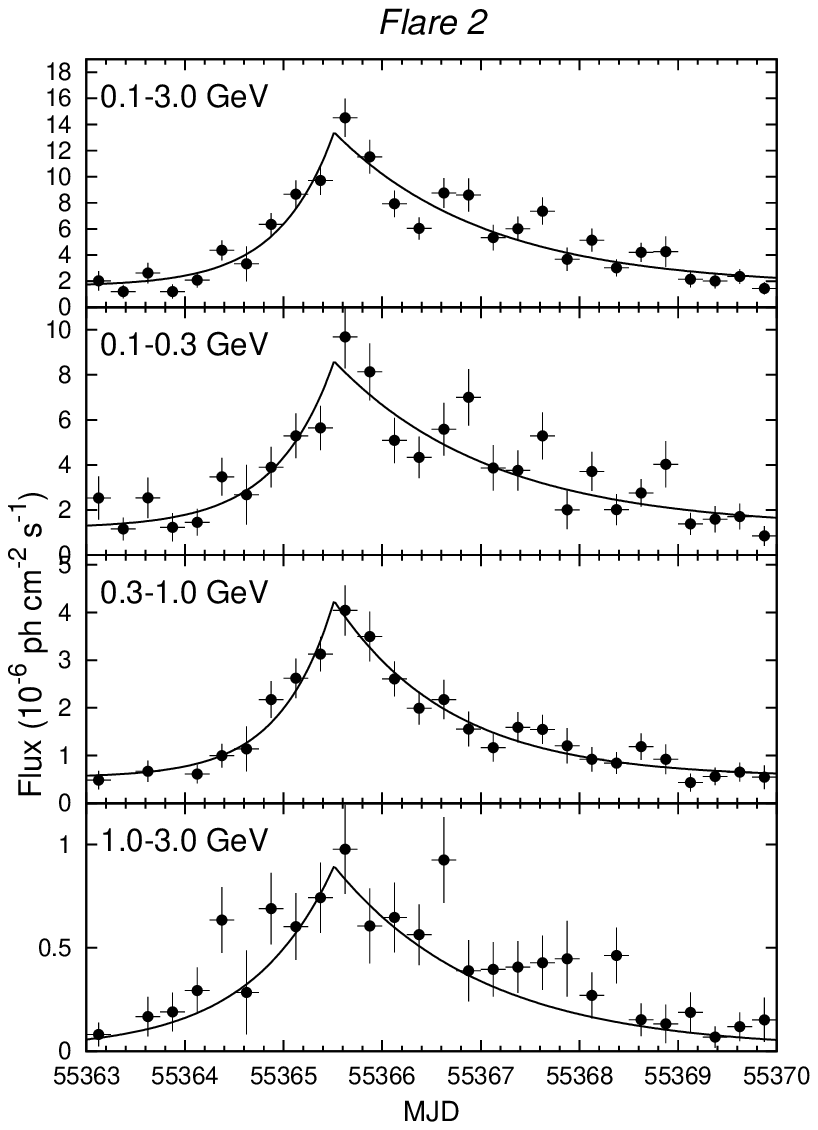}
\end{minipage}
\caption {6 hr $\gamma$-ray light-curves of PKS 1222+216 during \textit{Flare 1} (left) and
\textit{Flare 2} (right) in four different LAT energy bands. The solid black curves are the best
fit temporal profiles assuming an exponential rise and fall  (see Table \ref{tab:trtf}, \S\ref{subsec:var}).}
\label{fig:minipagelc}
\end{figure}

LAT $\gamma$-ray light-curves of PKS 1222+216 during \textit{Flare 1} and \textit{Flare 2}
in four different energy bands: 0.1 -- 3 GeV, 0.1 -- 0.3 GeV, 0.3 -- 1 GeV and 1 -- 3 GeV
at 6 hr timescale are shown respectively in the left and right panel of 
Figure \ref{fig:minipagelc}. Except for the 1--3 GeV emission, both flares have similar peak fluxes
in respective energy bands, but present a different case with respect to temporal features,
variability and sub-structures.

To access the asymmetry of rise and decay of the flares, we have fitted an exponential rising and falling profile
\begin{equation}
 f(t) = C+A(exp[(t-t_p)/t_r] + exp[-(t-t_p)/t_f]) \nonumber
\end{equation}
where $t_r$ and $t_f$ are rise and fall time in days with C, A and $t_p$ being a constant (quiescent flux),
normalization (half of peak flux) and the instant of the flare peak respectively. First, the quiescent flux
was derived by fitting a constant to a selected range of the 6 hr light-curve in each energy band. Fit 
was then performed on the 0.1 -- 3 GeV light-curves restricting the quiescent flux (C) to vary
within its best fit limits. The subsequent fitting in the other three energy bands was then performed by
fixing the peak position to this best fit value. The results
from the fits and associated $1\sigma$ errors are given in Table \ref{tab:trtf}. The fitted profiles, 
plotted only for data considered during the fitting, are shown in the left and right panel of Figure
\ref{fig:minipagelc}.

\begin{table}[ht]
\centering
\caption{Flare characteristics-- Best fit values with $1\sigma$ errors}
\begin{tabular}{c c c c c c c c}
\hline
Energy Band & No. of & C & A & $t_0$ & Rise time ($t_r$) & Fall time ($t_f$) & $\chi^2$ (dof) \\ 
\hline
(GeV) &	 time bins & \multicolumn{2}{c}{(10$^{-6}$ ph cm$^{-2}$ s$^{-1}$)} & (MJD) & (days) & (days) & \\
\hline \hline
 \multicolumn{8}{c}{\textit{Flare 1} }\\
\hline 
 0.1--3.0 & 20 & $1.5\pm0.2$  & $16.8\pm0.2$ & $55316.74\pm0.05$ & $0.6\pm0.1$ & $0.13\pm0.05$ & 30.9 (15) \\
 0.1--0.3 & 16 & $1.6\pm0.2$  & $9.3\pm1.6$  & --		  & $0.5\pm 0.1$ & $ 0.07\pm 0.02$ & 9.5 (12) \\
 0.3--1.0 & 18 & $2.3\pm0.2$  & $4.8\pm0.5$  & --		  & $0.9\pm 0.1$ & $ 0.20\pm 0.03$ & 19.7 (14) \\
 1.0--3.0 & 11 & $0.12\pm0.02$& $2.7\pm0.5$  & --		  & $0.4\pm 0.1$ & $ 0.11\pm 0.03$ & 11.3 (7) \\
\hline \hline
\multicolumn{8}{c}{\textit{Flare 2} }\\
\hline
0.1--3.0 & 28 & $1.8\pm0.2$  & $11.8\pm0.8$ & $55365.5\pm0.1$  & $0.6\pm0.1$ & $1.5\pm0.1$ & 49.2 (23) \\
0.1--0.3 & 28 & $1.4\pm0.2$  & $7.4\pm0.7$  & --		& $0.6\pm 0.1$ & $1.6\pm0.2$ & 34.4 (24) \\
0.3--1.0 & 26 & $0.57\pm0.03$& $3.7\pm0.3$  & --		& $0.5\pm 0.1$ & $1.2\pm 0.1$ & 15.2 (22) \\
1.0--3.0 & 27 & $1.1\pm0.2$  & $0.9\pm0.1$  & -- 		& $0.9\pm 0.2$ & $1.5\pm 0.2$ & 21.5 (23) \\
\hline 
\end{tabular}
\label{tab:trtf}
\end{table}

\subsection{Hardness Ratios}\label{subsec:hd}
\begin{figure}[!ht]
\centering
\begin{minipage}[t]{0.48\linewidth}
\includegraphics[scale=0.62,angle=0]{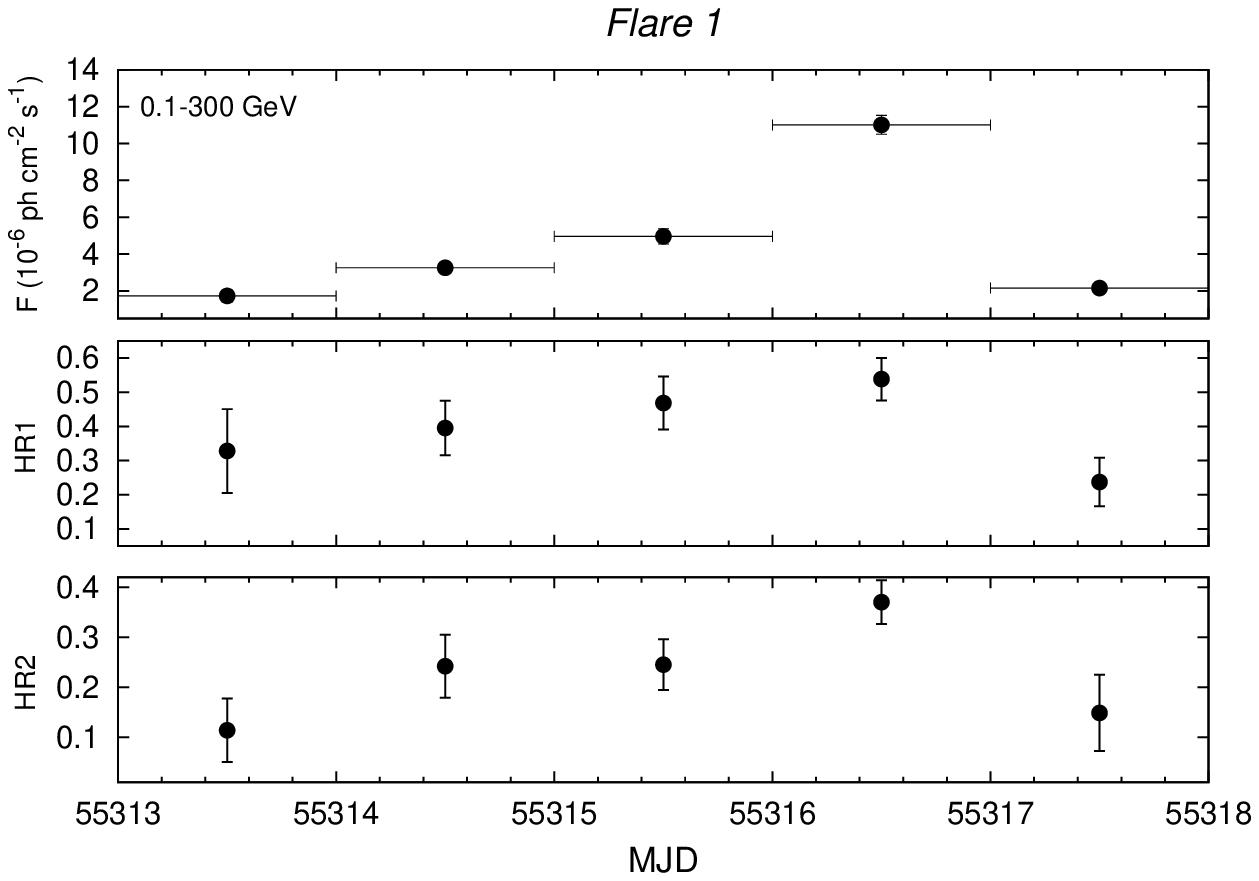}
\end{minipage}
 \quad
\centering
\begin{minipage}[t]{0.48\linewidth}
\includegraphics[scale=0.62,angle=0]{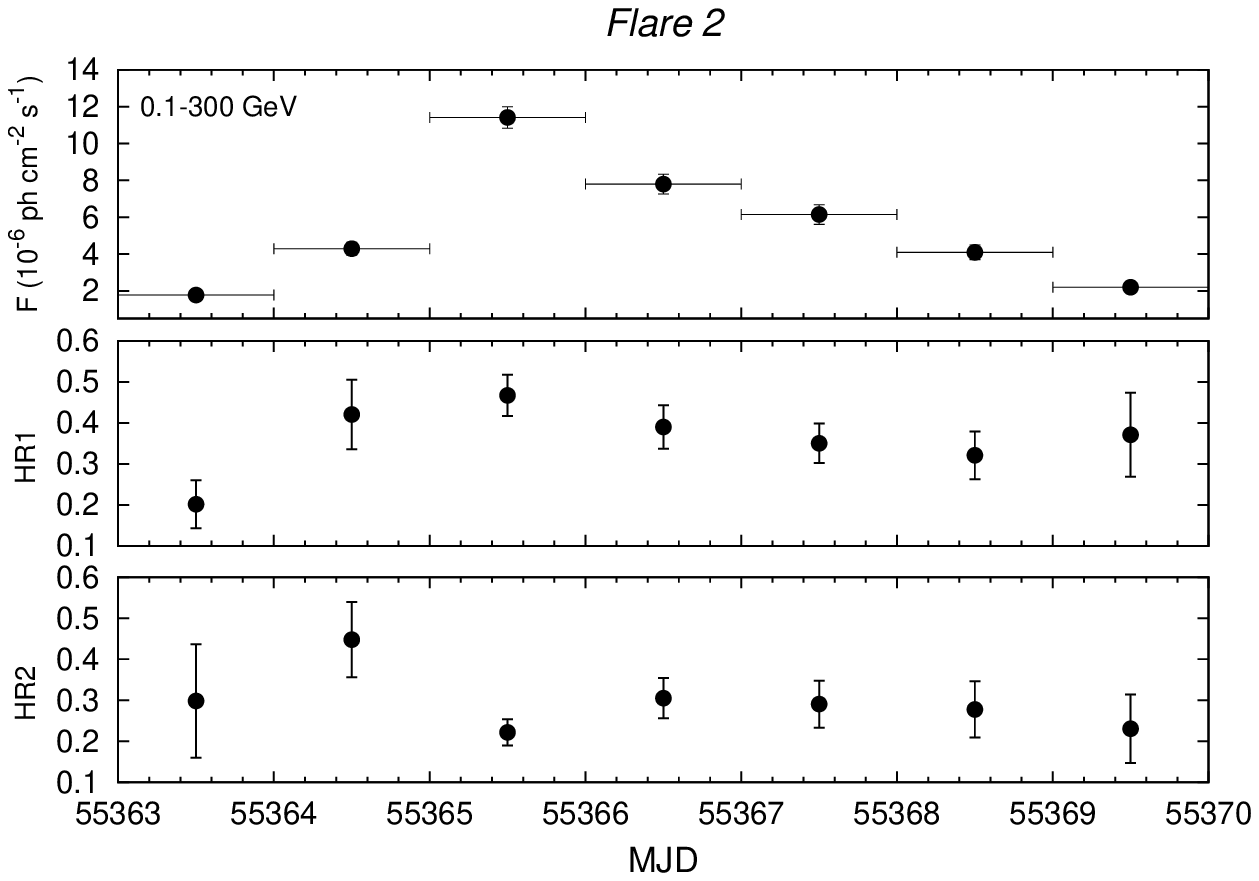}
\end{minipage}
\caption{Variability of flux (0.1--300 GeV) and Hardness ratios (HR1, HR2) on daily time-scale
during \textit{Flare 1} and \textit{Flare 2} (\S\ref{subsec:hd}).}
\label{fig:hd}
\end{figure}

The left and right panel of Fig. \ref{fig:hd} show the daily 0.1-300 GeV $\gamma$-ray
light-curve along with the corresponding hardness ratios (HRs) during \textit{Flare 1} and
\textit{Flare 2}. We have derived two hardness ratios, HR1 and HR2, defined as the ratio of
flux in 0.3--1 GeV to the corresponding flux in the 0.1--0.3 GeV band, and the ratio of 1--3 GeV
flux to the corresponding 0.3--1 GeV flux, respectively. Though
the peak fluxes are similar, the hardness ratios show very different evolution during the
rise, at the peak and during the decay phase of the flares.
Both HR1 and HR2, during \textit{Flare 1}, show a consistent hardening until the peak followed
by a softening to quiescent level during the flare decay. HR1 and HR2 during \textit{Flare 2},
on the other hand, suggest a hardening during the rising phase but a reverse trend at the peak
and a different evolution behavior during the decay phase. HR1 at peak is consistent with
hardening seen during the rising phase and softens during the decay but not to the quiescent
level. HR2, on the contrary, softens to quiescent level at the peak and remains at the same
level throughout the decay phase of \textit{Flare 2}.

\subsection{Time Delay/Lag} \label{subsec:lag}
It is interesting to note that VHE photons \citep{2010ATel.2617....1N, 2010ATel.2684....1M, 2011ApJ...730L...8A} 
precede the LAT peak during both the flaring episodes suggesting a soft lag\footnote{A clear peak in VHE
was detected during \textit{Flare 2} by the \textit{MAGIC} observatory on $\sim 55364.92$ but not during
\textit{Flare 1}}. In general, the origin of flares in BLR (torus) is expected to have absence (presence) of
time lag between the MeV and GeV emission \citep{2012ApJ...758L..15D}. Motivated by
this, we performed a lag analysis to look for a possible hint, if any, between the extracted 
LAT light-curves in different energy bands. The analysis is, however, limited by closely
separated energy bands and best available time resolution of 6 hr. The 
lag analysis was performed using \textit{z-transformed Discrete Correlation Function} (ZDCF) method of 
\citet{2013arXiv1302.1508A} (also see \citet{1997ASSL..218..163A}). It works on data pairs sorted according 
to their time lag and binned into equal population bins of at least 11 pairs after discarding multiple occurrences 
of the same data pair in a bin. Correlation coefficients of the bins are calculated and then z-transformed
to estimate the error in z-space which are then transformed back in the correlation space. DCF errors are
estimated using Monte Carlo simulations by adding a random error at each step to each data 
from the errors in the light-curves fluxes \citep[see][for more details]{2013arXiv1302.1508A}. 

\begin{figure}[!ht]
\centering
\begin{minipage}[t]{0.48\linewidth}
\includegraphics[scale=0.65,angle=0]{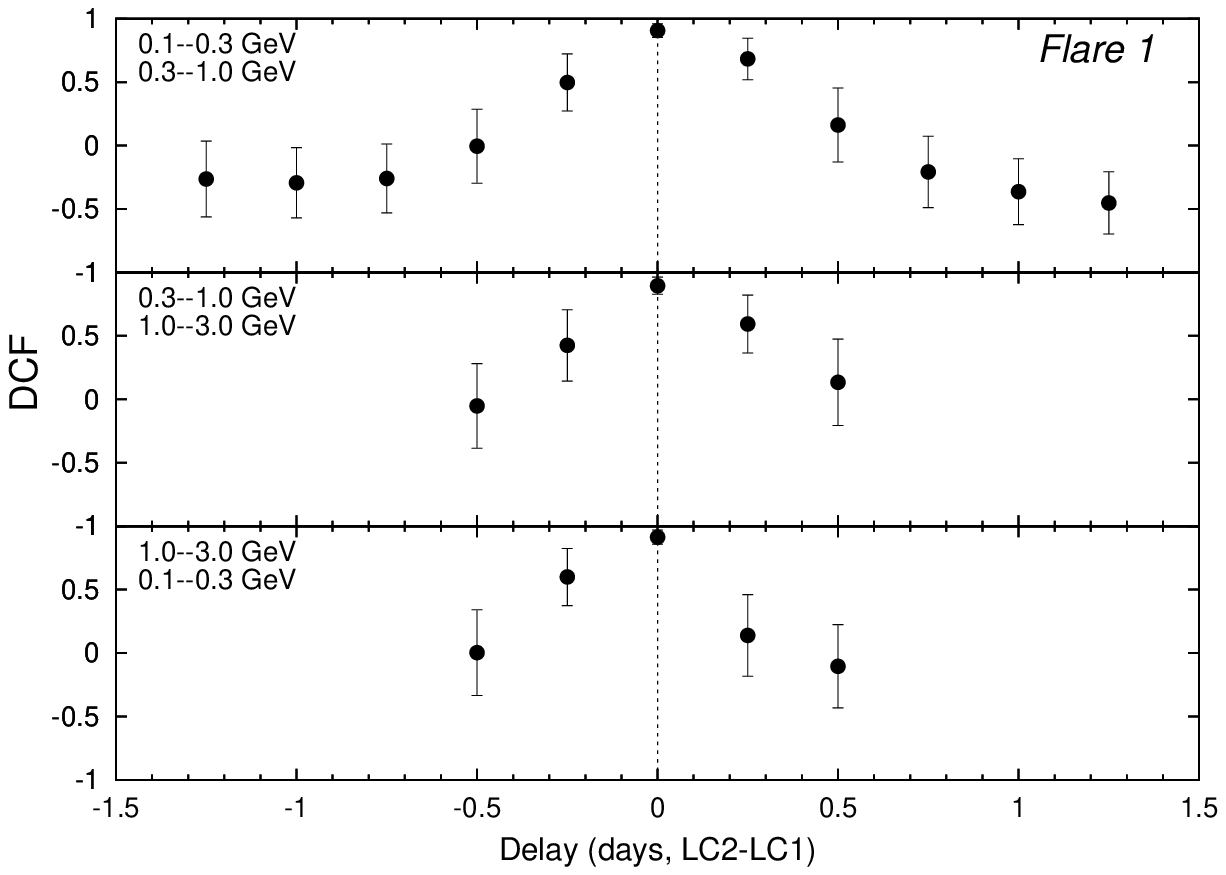}
\end{minipage}
 \quad
\centering
\begin{minipage}[t]{0.48\linewidth}
\includegraphics[scale=0.65,angle=0]{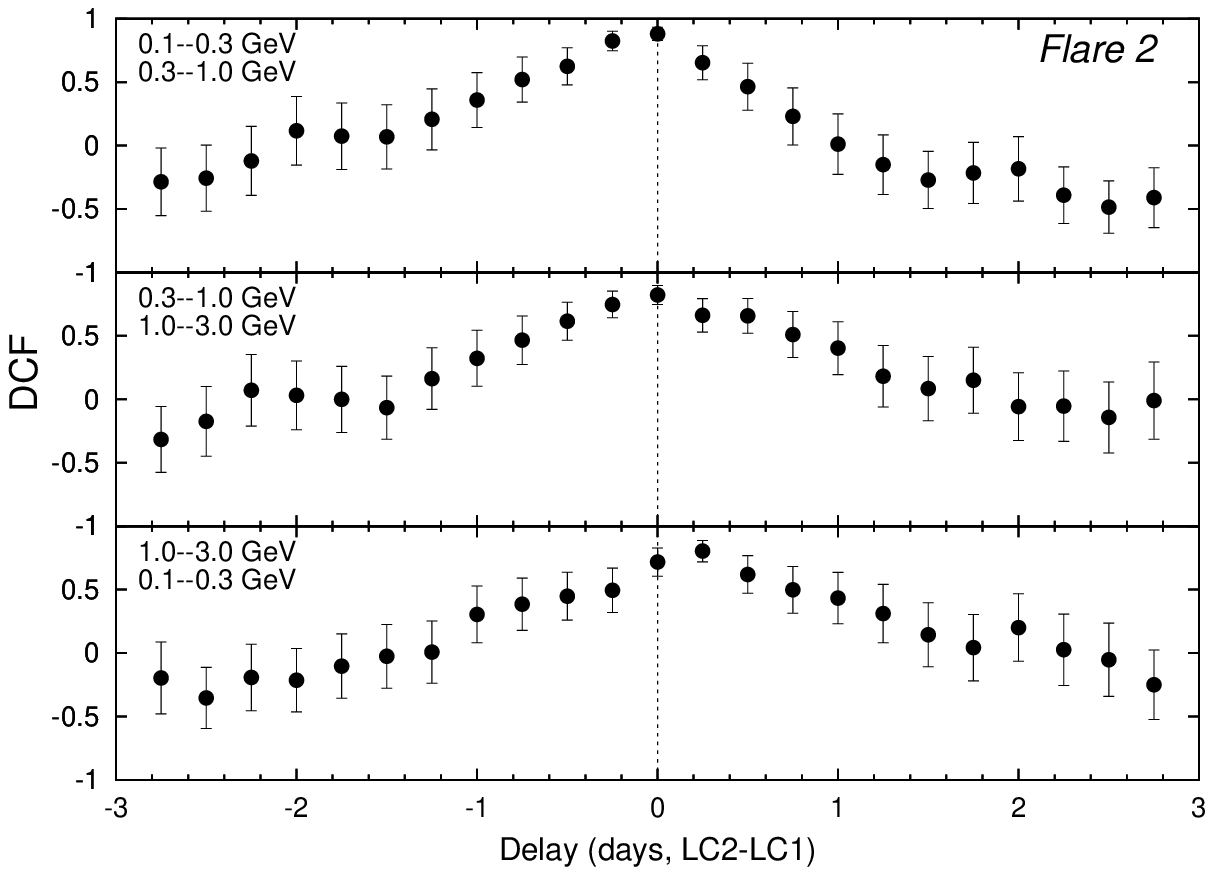}
\end{minipage}
\caption{DCF calculated using uniformly sampled 6 hr light-curves for \textit{Flare 1}
(left) and \textit{Flare 2} (right). The lag order used here is ``LC2--LC1'' light-curve
as mentioned in each panel (see \S\ref{subsec:lag}).}
\label{fig:lag}
\end{figure}

The time lag analysis was carried out on a continuous set of uniformly sampled data points during both
the flares, though the method used is also applicable in the case of non-uniformly sampled
light-curves\footnote{lag study using non-uniformly sampled 6 hr data is also consistent with quoted results}.
The results of the lag analysis derived from 1000 simulations of each pair, and the associated $1~\sigma$
uncertainties are given in Table \ref{tab:lag}. The corresponding DCFs for \textit{Flare 1} and \textit{Flare 2}
are presented in the left and right panel of Fig. \ref{fig:lag} respectively where the time ordering is ``T(2nd
light-curve:LC2) -- T(1st light-curve:LC1)''. Our analysis suggests a soft lag between the lowest and the
highest energy light-curves during \textit{Flare 2} while \textit{Flare 1} is consistent with no lag. Considering
the low significance of the inferred soft lag during \textit{Flare 2}, it has to be understood as an upper-limit only.
Further, there may be a lag among all the three light-curves but the limited time resolution provided
by the LAT light-curves and the closeness of energy bands makes it hard to ascertain.

\begin{table}[ht]
\centering
\caption{ Time lag$^\ast$ values obtained for {\it Flare 1} and {\it Flare 2} using uniformly sampled 6 hr LAT data}
\begin{tabular}{c c c c}
\hline \hline
 LC1 Energy Band & LC2 Energy Band & {\it Flare 1} lag & {\it Flare 2} lag	\\
 (GeV)		& (GeV)		  & (days)	     & ( days) \\
\hline
  0.1--0.3 & 0.3--1.0	&	0.0$^{+0.12}_{-0.11}$	&	0.0$^{+0.11}_{-0.19}$ \\
  0.3--1.0 & 1.0--3.0	&	0.0$^{+0.12}_{-0.11}$	&	0.0$^{+0.18}_{-0.21}$ \\
  1.0--3.0 & 0.1--0.3	&	0.0$^{+0.11}_{-0.12}$	&	0.25$^{+0.15}_{-0.20}$ \\
\hline
\multicolumn{3}{l}{$^\ast$ T(LC2) -- T(LC1) as given in columns 1 and 2}
\end{tabular}
\label{tab:lag}
\end{table}

\section{DISCUSSION}\label{sec:discussion}
In this work, we have performed a detailed study of the two brightest $\gamma$-ray flares
of PKS 1222+216 observed in \textit{Fermi}-LAT in April and June 2010. The daily 0.1--300 GeV
light-curve during these two flares has been systematically studied by dividing it into
multiple energy bands (0.1--3, 0.1--0.3, 0.3--1 and 1--3 GeV) and using the shortest
possible time binning (6 hr) allowed by the photon statistics. Contrary to a smooth rise
followed by a rapid (\textit{Flare 1}) or gradual (\textit{Flare 2}) fall
observed in the daily 0.1 -- 300 GeV LAT $\gamma$-ray light-curves (top panel, Fig. 3), 
the 6 hr light-curves present a different case (Fig. \ref{fig:minipagelc}). These 6 hr
light-curves reveal a complex temporal behavior and variability patterns in different energy
bands during the flaring episodes. \textit{Flare 1} begins with a $\sim 2$ day long plateau
in the 0.1--0.3 GeV light-curve and enhanced variations at high energies with an apparent 
daily recurring pattern in 1--3 GeV light-curve. After this phase, a sharp increase in the flux 
is observed in all the LAT bands, followed by a rapid descent. To our knowledge, a plateau phase
before flare has been a characteristic property of blazar 3C\,454.3 \citep{2011ApJ...733L..26A}
but has not been seen in any other blazar during a flare. The absence of the plateau phase
in the total energy integrated ($0.1-300$ GeV) daily light curve during \textit{Flare 1},  suggests
the rise may be governed primarily by $>0.3$ GeV photons (top panel, Fig. \ref{fig:hd}).
In addition, presence of a plateau phase at lower energies is also reflected as the slow
rise of daily binned 0.1-300 GeV $\gamma$-ray light curve during \textit{Flare 1} (Figs.
\ref{fig:lcspec} \& \ref{fig:hd}--top panel, \S3). An exponential fit during the rising phase of the time 
and energy resolved light curves indicate a similar rise in all the bands except 0.3--1 GeV
(Table \ref{tab:trtf}, Fig. \ref{fig:minipagelc}). If we consider a situation where a short
lived acceleration mechanism (or a shock) is initiated in a turbulent jet,
then the plateau phase may correspond to a case where the jet is becoming
turbulent followed by the sudden rise probably due to a short lived acceleration. When
the jet becomes turbulent, the electrons are accelerated via second order Fermi
mechanism on a characteristic timescale, $t_{\rm acc}^{\rm (t)} \sim (r_{\rm g}/c)
(c/V_{\rm A})^2$, with acceleration rate $\propto 1/t_{\rm acc}$
\citep{2007Ap&SS.309..119R, 2004PASA...21....1P, 2005NewA...11...17B}. Here, $r_{\rm g}$
is the electron gyro-radius and $V_{\rm A}$ is the Alfven speed. On the other hand, shock
acceleration is a first order Fermi mechanism with acceleration timescale
$t_{\rm acc}^{\rm (s)} \sim (r_{\rm g}/c) (c/u_{\rm s})^2$, where $u_{\rm s}$ is the speed
of shock \citep{2001JPhG...27.1589K, 2004PASA...21....1P, 2007Ap&SS.309..119R}. Hence, the
turbulent acceleration rate lags the shock acceleration by a factor $\sim(V_{\rm A}/u_{\rm s})^2$
and the latter dominates the emission mechanism, once initiated.

In the case of \textit{Flare 2}, the time and energy resolved light curves show a monotonic
exponential rise from the beginning followed by a gradual descent with signatures
of multiple peaks of decreasing strength from lower to higher energies (Fig. \ref{fig:minipagelc}).
The rate of rise of flux in all the energy bands are similar and also consistent with that of
\textit{Flare 1} (Table \ref{tab:trtf}). The peak fluxes, during both the flares, are also
similar in the respective energy bands except in the 1--3 GeV band during \textit{Flare 1},
where the peak reaches approximately twice the corresponding peak flux during the \textit{Flare 2}. This is,
however, consistent with the observed ``harder when brighter'' trend seen during \textit{Flare 1}
(right panel, Fig. \ref{fig:lcspec}). Though the rate of rise are similar, the monotonic
rise, lacking a plateau phase, probably causes the faster rise of the total energy integrated
($0.1-300$ GeV) daily-light curve compared to \textit{Flare 1}. On the other hand, the slow decline of
the daily, energy integrated 0.1--300 GeV light-curve is mainly due to the dominant
contribution of 0.1--0.3 GeV emission with multiple peaks structures. This may be due to the
contribution from multiple emission regions and their relative locations \citep{2013MNRAS.430.1324N,
2013MNRAS.431..355G} in addition to the standard flaring region contributing across the spectrum.

\begin{figure}[!ht]
 \centering
  \includegraphics[scale=1.0]{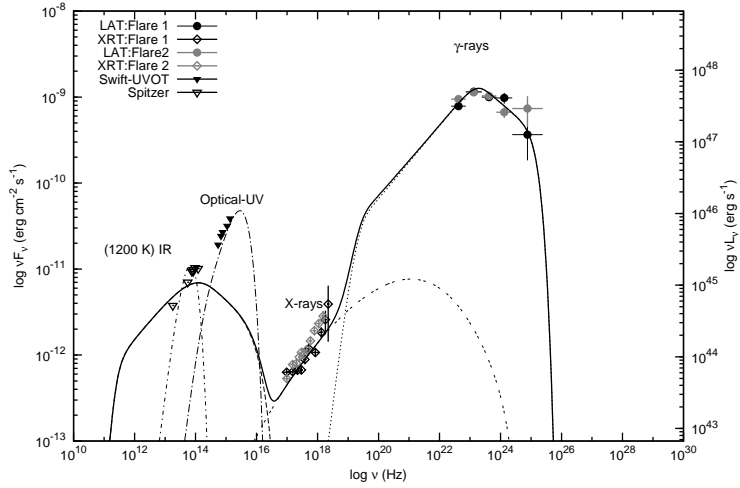}  
  \caption{Broadband SED of PKS 1222+216 during \textit{Flare 1} and \textit{Flare 2} along with
  the model spectrum (\S4).}
  \label{fig:SED}
\end{figure}

In contrast to the rising phase, the descent of the flux is faster in case of \textit{Flare 1}
but rather gradual in case of \textit{Flare 2} (see Section \S3). To further investigate the
flare decay we obtained the physical parameters of the source by modeling the broadband SED
using one zone leptonic model \citep{2013MNRAS.433.2380K,2012MNRAS.419.1660S}. We used 
simultaneous/contemporaneous time averaged X-ray and LAT spectrum over the LAT flaring duration. The 
X-ray emission is reproduced considering SSC process whereas the $\gamma$-ray spectrum is
reproduced considering external Compton scattering of the IR photons from a 1200 K dusty torus. 
The observed fluxes in X-rays and $\gamma$-rays during both the flares are
similar and hence we assume that same set of physical parameters describe the
source SED during the flares. The size of the emission region is constrained via the fastest
observed rise time ($0.4$ day) while the particle indices are deduced from the X-ray and $\gamma$-ray spectra
\citep{2013MNRAS.433.2380K}. The final model spectrum along with the observed fluxes are shown
in Fig. \ref{fig:SED} and the corresponding parameters are given in Table \ref{tab:SEDpar}. Based
on these parameters, one can estimate the cooling timescale of the electrons emitting $\gamma$-rays
in the observer's frame as \citep{2013MNRAS.433.2380K,2014MNRAS.442..131K,2013ApJ...766L..11S}
\begin{align}\label{eq:cool}
 t_{\rm cool} &\simeq (3m_{\rm e} c/4 \sigma_{\rm T} U_{\rm IR}^\prime)\times \sqrt{(1+z) \epsilon_\ast/\epsilon_{\rm\gamma}} \nonumber \\
	  & \sim 4 \left(\frac{\xi_{\rm ir}}{0.15}\right)^{-1}\left(\frac{\Gamma}{22}\right)^{-2} 
	  \left(\frac{T_\ast}{1200 K}\right)^{-7/2}\left(\frac{\epsilon_{\rm\gamma}}{2 ~GeV}\right)^{-1/2} \text{min}
\end{align}
where $\xi_{\rm ir}$ ($\sim L_{\rm IR}/L_{\rm UV}$) is the IR covering fraction with $L_{\rm UV}$
as the disk luminosity \citep{2013MNRAS.433.2380K}, m$_{\rm e}$ is the rest mass
of an electron, c is the speed of light and $\sigma_{\rm T}$ is the Thomson scattering
cross-section. The IR photon energy density in the emission frame, $U_{\rm IR}^\prime$
is related to the corresponding AGN frame energy density ($U_{\rm IR}$) as $U_{\rm IR}^\prime = \Gamma^2\, U_{\rm IR}$. 
The estimated cooling time scale (equation \ref{eq:cool}) is much smaller than the
decay time of the flares obtained through the exponential fit of the light curves
(Table \ref{tab:trtf}) and hence we can conclude that decay of both the flares cannot
be attributed to radiative cooling processes alone. If the decline is due to light travel time effects
then one expects a similar decay times in all the energy bands, contrary to the one obtained
(see Table \ref{tab:trtf}).
Alternatively, the flare decay can also be associated with the weakening of the acceleration
process. In this case, one expects a steepening of spectral indices \citep{1998A&A...333..452K}.
Both HR1 and HR2 during \textit{Flare 1} show hardening during the rise followed by a
softening to quiescent level (left panel, Fig. \ref{fig:hd}) during the decay. Relating this feature
to the efficiency of particle acceleration process, we interpret the hardening of spectrum during
rise as growth of shock acceleration process followed by its weakening, thereby softening the spectra
during the flare decay. Contrary to the trend seen in \textit{Flare 1}, the hardness ratios of \textit{Flare 2} 
exhibit complex behavior. Here HR1 seems to harden till peak followed by a milder softening during decay
whereas HR2 softens to its quiescent level at the peak and remains consistent with
it throughout the decay (right panel, Fig. \ref{fig:hd}).
Hence, unlike \textit{Flare 1}, \textit{Flare 2} cannot be interpreted as a result of single
acceleration process but probably include other dynamical effects. The decay of \textit{Flare 2}
on daily timescales has been explained satisfactorily by considering the effects of jet
dynamics. A decelerating jet interpretation can successfully reproduce the overall
observed light-curves and SED \citep{2014MNRAS.442..131K}, though it seems that this 
model alone is not sufficient to explain the  observed complex temporal behavior presented
here. Nevertheless, this interpretation is supported by non-detection of significant
lags within various energy bands during the flare. The $\sim6$ hr lag observed in
high energy during \textit{Flare 2} can be a manifestation of different cooling processes
\citep{2012ApJ...758L..15D} and/or blob dynamics. Similar lags have also been found during
the bright flares of FSRQ PKS 1510-089 \citep{2013MNRAS.431..824B}. However, the inferred
lag is hard one contrary to soft lag found in our study. The absence of lag during
\textit{Flare 1} might be due to shorter duration of significant variation and
averaging of data in the analysis of the same (see \S3.2).

\begin{table}[!ht]
 \centering
 \begin{minipage}[t]{0.50\textwidth}
 \caption{SED parameters for \textit{Flare 1} and \textit{Flare 2}}
 \begin{tabular}{l c} 
 \hline
  Parameters & Numerical values (CGS units) \\ \hline
  Particle index before break (p)		& 2.15	\\
  Particle index after break (q)		& 3.6	\\
  Magnetic field (B)			& 0.35	G\\
  Equipartition factor ($\eta$)		& 40 \\
  Doppler factor ($\delta$)		& 23	\\
  Angle to the line of sight ($\theta$)	& 2.5$^\circ$	\\
  Particle break energy$^\ast$ ($\gamma_{\rm b^\prime}$)	& $2.3 \times 10^3$	\\
  Emission region size (R$^\prime$)	& $1 \times 10^{16}$	\\
  IR Torus temperature (T$_\ast$)		& 1200 K \\
  Jet power ($P_{\rm jet}$)			& $2\times 10^{46}$	\\
  Minimum particle energy$^\ast$ ($\gamma_{\rm min}^\prime$) & 25	\\
  Maximum particle energy$^\ast$ ($\gamma_{\rm max}^\prime$) & $2 \times 10^4$	\\
 \hline
 $^\ast$in electron rest-mass units
  \end{tabular}
  \end{minipage}
 \label{tab:SEDpar}
\end{table}

Based on lag and correlated-variability, both homogeneous and inhomogeneous model have
been suggested for blazars' emission and have been successful in reproducing the observed
SED and light-curves of the sources \citep{2011A&A...534A..86T, 2013MNRAS.431..824B,
2013MNRAS.433.2380K, 2014MNRAS.442..131K}. Simple one-zone models have been found to satisfactorily
reproduce the broadband SED during both the flares \citep{2011A&A...534A..86T,
2014ApJ...786..157A,2014MNRAS.442..131K}. However, the rise and fall within a day, during 
\textit{Flare 1}, suggests that it probably arises from a $\sim0.5$ day long emission
region as suggested by a similar rise time in all the LAT energy bands\footnote{This is different from
the $\sim10$ minutes variability considered by \citet{2014ApJ...786..157A} based on the VHE
variability observed during \textit{Flare 2}}. An indication of this daily variation is also apparent
in the 1--3 GeV emission before rise. Moreover, contrary to this variation
in 1--3 GeV emission, the 0.3--1 GeV emission suggests faster variation and has slowest rise
and fall compared to the other two energy bands (Table \ref{tab:trtf}).
For \textit{Flare 2}, various models including leptonic scenarios e.g.
jet-in-jet/minijet \citep{2011A&A...534A..86T,2013MNRAS.431..355G}, recollimation 
\citep{2011ApJ...730L...8A, 2011A&A...534A..86T, 2012MNRAS.425.2519N, 2014MNRAS.442..131K}
and multiple emission zones \citep{2011A&A...534A..86T} along with hadronic scenario
of ultra-relativistic neutral beams \citep{2012ApJ...755..147D} have been proposed.
However, the temporal features seen in the systematic time and energy resolved study
performed here demand scenarios beyond simple one zone homogeneous models where minor 
fluctuations are embedded within the region contributing to the major flare 
\citep{2012A&A...539A.149H,2013MNRAS.431..355G}.

The two flares considered here have been studied earlier by \citet{2011ApJ...733...19T} and
\citet{2013MNRAS.430.1324N}. Both the previous studies focus on the spectral and temporal
features in the 0.1--300 GeV energy band employing different time binning method. In
this work, however, we have a performed a more detailed and systematic time and energy
resolved study. We have shown that all the features/patterns observed during \textit{Flare 2} by 
\citet{2013MNRAS.430.1324N} are mainly due to features present at lower energies. \citet{2011ApJ...733...19T},
on the other hand have termed these two flares as the highest possible
$\gamma$-ray fluxes from PKS 1222+216, supposedly fueled by feeding the entire accretion
power to the jet. However, the observed peak luminosity ($L_{obs}$) which reaches $\sim 10^{48}$ erg s$^{-1}$
during the peak corresponds to an emitted power of $L_{em}\sim L_{obs}/2\Gamma^2 \simeq
1\times 10^{45}$ erg s$^{-1}$ \citep{1997ApJ...484..108S} in the rest frame of the emission region,
assuming a bulk Lorentz factor ($\Gamma$) of 20 \citep{2012IJMPS...8..163H, 2014ApJ...786..157A}. This is
$\sim10\%$ of the accretion disk luminosity $L_d \sim 3.5\times 10^{46}$ erg s$^{-1}$ \citep{2011ApJ...732..116M}
and is in contrast to $L_{em} \sim L_d $ derived by \citet{2011ApJ...733...19T}, who assumed a
comparatively lower value for the accretion disk luminosity $L_d\simeq 5 \times10^{45}$ erg s$^{-1}$
(assuming a BLR covering fraction $\xi_{BLR} \simeq 0.1$) and jet bulk Lorentz factor $\Gamma \simeq
10$. A black hole of mass $\sim 6 \times 10^8 M_\odot$ \citep{2012MNRAS.424..393F}, however, can
produce an Eddington luminosity of $\sim 8 \times 10^{46}$ erg s$^{-1}$ which in turn suggests a very high 
accretion power and disk radiative efficiency. Apart from the asymmetry observed in the light-curves
of the flares, \citet{2011ApJ...733...19T} found a hint of ``harder-when-brighter'' during \textit{Flare 1}
and a clockwise evolution in spectral index vs flux during \textit{Flare 2}. Our analysis also
reproduces these results but we found a clockwise trend during \textit{Flare 1} as well,
which is present in the study of \citet{2011ApJ...733...19T} but was probably suppressed 
due to the inclusion of a flare preceding the \textit{Flare 1}.

\section{CONCLUSIONS}\label{sec:conclude}
We have performed an in-depth time and energy resolved study of the two brightest $\gamma$-ray
flares from FSRQ PKS 1222+216 observed by the \textit{Fermi}-LAT during April (\textit{Flare 1})
and June (\textit{Flare 2}) 2010. Our study reveals a large variety of temporal features and 
variability patterns in different energy bands (0.1--3, 0.1--0.3, 0.3-1.0 and 1--3 GeV) binned
on 6 hr timescales apart from a clearly asymmetric profile in both the flares. This includes
a $\sim 2$ day plateau in 0.1--0.3 GeV band, hint of variation on daily timescale in 1--3 GeV
emission, faster fluctuation ($\sim 0.5$ days) at 0.3--1 GeV energy
and a rapid decline during \textit{Flare 1}. \textit{Flare 2}, on the contrary, shows a monotonic rise
followed by a gradual decline at all energies with prominent substructures in 0.1--0.3 GeV emission.
Though the rise time are similar in all energy bands during both the flares, the slower rise of daily
integrated 0.1--300 GeV light-curve during \textit{Flare 1} is probably due to the presence of plateau
in the 0.1--0.3 GeV emission with $>0.3$ GeV photons driving the initial rise. Further, the slower decline
of 0.1--300 GeV light-curve during \textit{Flare 2} is mainly due to contribution from the 
multiple peaked sub-structures in the 0.1--0.3 GeV light-curve resulting in a coherent single flare at
daily timescales. 

SED during both the flares can be well reproduced by a simple one zone model considering synchrotron,
SSC and EC of IR emission mechanisms with similar parameters. The radiative cooling timescale of
$\sim 4$ min suggest that the observed duration and profiles of flares are not solely due to radiative
cooling. The decline of flares also cannot be explained by considering light travel time effect which
results in a similar decline rate at all energies, contrary to observations. 

Study of hardness ratios
during both the flares suggest that \textit{Flare 1} can result from a variation in the
efficiency of underlying acceleration process whereas for \textit{Flare 2}, one need to
consider the effect of jet dynamics and small scale inhomogeneities. The former contribute to overall 
flaring while the latter is responsible for the intrinsic feature observed in the time and energy 
resolved light-curves presented here. Based on these features we suggest that both flares
are probably result of two different underlying mechanisms that are equally efficient in producing 
luminous $\gamma$-ray flares.

The \textit{Fermi}-LAT Collaboration acknowledges generous ongoing support from a number of agencies and 
institutes that have supported both the development and the operation of the LAT as well as scientific
data analysis. These include the National Aeronautics and Space Administration and the Department of 
Energy in the United States, the Commissariat $\grave{a}$ l'Energie Atomique and the Centre National de la Recherche
Scientifique/Institut National de Physique Nucl$\grave{e}$aire et de Physique des Particules in France, the
Agenzia Spaziale Italiana and the Istituto Nazionale di Fisica Nucleare in Italy, the Ministry of Education,
Culture, Sports, Science and Technology (MEXT), High Energy Accelerator Research Organization (KEK) and Japan
Aeros pace Exploration Agency (JAXA) in Japan, and the K. A. Wallenberg Foundation, the Swedish Research Council
and the Swedish National Space Board in Sweden. 

Additional support for science analysis during the operations phase from the following agencies is a
lso gratefully acknowledged: the Istituto Nazionale di Astrofisica in Italy and
and the Centre National d'Etudes Spatiales in France.

\end{document}